\documentclass[preprint,superscriptaddress]{revtex4}
\usepackage{graphics,epsfig}
\usepackage{epstopdf}
\usepackage{graphicx}
\usepackage{dcolumn}
\usepackage{amsmath}
\usepackage{hyperref}
\newcommand{\be}{\begin{equation}}
\newcommand{\ee}{\end{equation}}
\newcommand{\bea}{\begin{eqnarray}}
\newcommand{\eea}{\end{eqnarray}}
\newcommand{\nn}{\nonumber}
\def\be{\begin{equation}}
\def\ee{\end{equation}}
\def\ba{\begin{eqnarray}}
\def\ea{\end{eqnarray}}
\begin{document}
\title{ Extended phase space thermodynamics for  Bardeen black holes  in massive gravity}

\author{ Benoy Kumar Singh}
\email{bksingh100@yahoo.com}
\author{Raj Pal Singh}
\email{rpsinghmathura@gmail.com}
\author{Dharm Veer Singh}
\email{veerdsingh@gmail.com}
\affiliation{Department of Physics, Institute of Applied science and Humanities, GLA University, Mathura 281406, India}

\begin{abstract}
 \noindent  This paper presents an exact solution of Bardeen black hole in  presence of massive gravity,  which is characterized by the additional parameter $m$. Here, we focus on $4D$ Bardeen AdS  massive black hole solutions as example to discuss the thermodynamical properties such as temperature, entropy and specific heat at constant pressure. We  also study the critical behaviour of Bardeen AdS massive black holes  by considering the cosmological constant $\Lambda$ as thermodynamical variable (pressure) as well as the parameter associated with the nonlinear electrodynamics. We calculate the critical value of pressure and temperature and study the effect of magnetic charge $e$ and mass parameter $m$.   It is  seen  that the thermodynamical volume are the independent of  mass parameter $m$.  The critical values are the one in which phase transition takes place and the nature of mass parameter $m$ and magnetic charge $e$ are opposite to each other. The critical temperature and pressure  were highly sensitive for these  parameters. 
\end{abstract}

\maketitle

\section{Introduction}
\noindent Black holes, coined by Wheeler \cite{wh67} were generally considered as rather esoteric objects of purely theoretical interest and little physical relevance. It is well known that  a black hole is thermodynamical system, there exist fundamental connections between the thermodynamics and general relativity since the  discovery of black hole entropy by Bekenstein \cite{JD,JD1,JD2,JD3,JD4}  and Hawking \cite{SWH}. During the past years the development of Maldasena conjucture (AdS/CFT correspondence) \cite{jm}, has attracted significant attention to AdS black holes.  The properties of  black hole physics have changed dramatically and subsequent studies found that the AdS black hole behave like the Van der Walls fluid. 

The first regular black hole was proposed by Bardeen \cite{Regular:1968} which means there is no singularity. Later on an exact solution of  Bardeen black hole was given by Ayon-Beato and Garcia \cite{AGB,AGB1,AGB2} wherein general relativity coupled to nonlinear electrodynamics. Thereafter, more intense emphasis was given by researchers investing the regular black holes and mor regular black holes were discovered recently \cite{hc,lbev,Balart:2014cga,Xiang,Bronnikov:2000vy,Zaslavskii:2009kp,Lemos:2011dq,Ansoldi:2008jw,Ghosh:2014pba,dvs99}. But most of these solutions are based on Bardeen's Model \cite{Regular:1968}. In later years the generalized solutions of Bardeen black hole model were developed which includes Bardeen de Sitter black hole  \cite{singh,fr1}, rotating  Bardeen solution \cite{Bambi},  noncommutative Bardeen solution \cite{sharif}, higher dimensional black holes  \cite{sabir} and EGB black holes \cite{dvs19,kumar19,Kumar:2020bqf,Singh:2020xju,Kumar:2020uyz} etc. However,  many singular black hole solutions with massive gravity  \cite{Cai:2014znn,Babichev:2015xha,EslamPanah:2019fci,EslamPanah:2018rob} are already present in the literature, but they  are not regular black hole solutions. Therefore, the aim of this work to get $4\-D$  spherically symmetric  Bardeen-like black hole solution in massive gravity  in AdS spacetimes viz., Bardeen AdS massive metric.  It is  shown that the Bardeen AdS massive metric is an exact black  hole solution  in AdS spacetime thereby generalizing the  Bardeen black hole solution \cite{Tzikas:2018cvs} which is encompassed as a special case. We analyze  their thermodynamical properties  and also perform a  phase structure  analysis of the Bardeen AdS  massive black holes.

In the extended phase space cosmological constant ($\Lambda$) is identified as thermodynamical  pressure ($P=-\Lambda/8\pi$) and its conjugate quantity is  thermodynamical volume \cite{Cvetic:2010jb, dolan13}.  Hawking and Page \cite {hp} have first studied the phase transition between the AdS black hole and thermal AdS space  and also confinement/deconfinement phase of gauge field  was studied by Witten \cite{witten}. The phase transition behaviour of charged AdS black hole and  the Van der Walls liquid-gas system were studied by Chamblin \cite{Chamblin:1999tk,Chamblin:1999hg}. In fact, phase transition plays an important role to investigate the thermodynamical properties of the objects at the critical point.  This consideration  has been investigated for different type   of black holes  \cite{1,2,3,4,5,6,7,8,9,10,11,12,13,14,15,16,17,18,19,20,21,22,23,24}.

The paper is organized as follows. In Section II we find the exact solution of Bardeen black hole in the presence of massive gravity and studied its physical and thermodynamical properties.  Section III,  gives the investigation of  phase structure for  Bardeen AdS  black holes in massive gravity. The thermodynamical stability and phase diagrams are studied in Sec IV.   Finally, we discuss our result and conclusions in Section V. (Here we use the units
$G_4 =  k_B = c = 1$).
\section{\label{sec:level3}Bardeen Black Hole solution in Massive gravity}
The Einstein- Hilbert action coupled to nonlinear electrodynamics in the presence of massive gravity  with negative cosmological constant \cite{cao} is given by
\begin{eqnarray}
S &=&\frac{1}{2 }\int d^{4}x\sqrt{-g}\Big[R
-2\Lambda +m^{2}\sum_{i}^{2}c_{i}\,\mathcal{U}_{i}(g,f)- \frac{1}{4\pi}\mathcal{ L}(F)\Big],
\label{action}
\end{eqnarray}
where $R$ is scalar curvature, $\Lambda$ is a cosmological constant,  $m$ is a parameter of massive gravity,  $c_i$ are constants, $f$ is symmetric tensor and $\mathcal{U}_{i}$ are polynomials of eigenvalues of the ($4\times 4$) matrix $\mathcal{K}_{\nu }^{\mu }=\sqrt{g^{\mu \alpha }f_{\alpha \nu }}$ which can be written as   
\begin{eqnarray}
\mathcal{U}_{1} &=&\left[ \mathcal{K}\right] ,\nn \\
\mathcal{U}_{2} &=&\left[ \mathcal{K}\right] ^{2}-\left[ \mathcal{K}^{2}
\right] ,\nn\\
\mathcal{U}_{3} &=&\left[ \mathcal{K}\right] ^{3}-3\left[ \mathcal{K}\right] 
\left[ \mathcal{K}^{2}\right] +2\left[ \mathcal{K}^{3}\right] ,\nn \\
\mathcal{U}_{4} &=&\left[ \mathcal{K}\right] ^{4}-6\left[ \mathcal{K}^{2}
\right] \left[ \mathcal{K}\right] ^{2}+8\left[ \mathcal{K}^{3}\right] \left[
\mathcal{K}\right] +3\left[ \mathcal{K}^{2}\right] ^{2}-6\left[ \mathcal{K}
^{4}\right],
\end{eqnarray}
with $\left[ \mathcal{K}\right]=\left[ \mathcal{K}_{\mu }^{\mu }\right]$ and $\mathcal{ L}(F)$ is the function of $F\equiv F_{\mu \nu } F^{\mu \nu}$   be  the electromagnetic field  tensor  which is  the generalization of Maxwell field and it reduces to Maxwell field in the weak field limit. We consider Lagrangian density of  the non-linear electromagnetic field as \cite{AGB2}
\begin{eqnarray}
{\cal{L}}(F)=\frac{1}{2se^2}\left(\frac{\sqrt{2e^2F}}{1+\sqrt{2e^2F}}\right)^{\frac{5}{2}},
\label{matter}
\end{eqnarray}
where $\mathcal{ L}(F)$ is the Lagrangian density of Bardeen source \cite{AGB2}  and parameters  $M$ and  $e$ connected with $s$ by relation $s=e/2M$. We would like to consider the spherically-symmetric  static metric of the spacetime, given by the following  line element
\begin{equation}
ds^2=-f(r)dt^2 +\frac{1}{f(r)}dr^2+r^2d\Omega^2, \qquad\text{with}\qquad f(r)=1-\frac{2m(r)}{r}.
\label{m2}
\end{equation}
We consider the following metric ansatz \cite{cao} for the reference metric as
\begin{equation}
f_{\mu \nu }=diag(0,0,c^{2}h_{ij}),
  \label{f11}
\end{equation}
where $c$ is  the positive constant. Using the metric ansatz (\ref{f11}), one can easily write $\mathcal {U}_i$ as
\begin{eqnarray}
&&\mathcal {U}_1=\frac{2}{r},\qquad \mathcal {U}_2=\frac{2}{r^2} ,\qquad \mathcal {U}_3=0 \quad\text{and}\quad \mathcal {U}_4=0.
\end{eqnarray}
Using the action (\ref{action}) and variation of this action with respect to the metric tensor ($g_{\mu \nu }$) and the electromagnetic  potential ($A_{\mu}$), respectively, leads to
\begin{eqnarray}
G_{\mu \nu }&+&\Lambda g_{\mu \nu }+m^{2}\chi _{\mu \nu }=T_{\mu\nu}  \equiv2\left[\frac{\partial {\cal{L}}(F)}{\partial F}F_{\mu\rho}F^{\rho}_{\nu}-g_{\mu\nu}{\cal{L}}(F)\right],\nn\\
&&\nabla_{\mu}\left(\frac{\partial {\cal{L(F)}}}{\partial F}F^{\mu \nu}\right)=0,\quad \text{and}\quad \nabla_{\mu}(* F^{\mu\nu})=0,
\label{Field equation}
\end{eqnarray}
 where $T_{\mu\nu}$ is the energy momentum tensor, $G_{\mu\nu}$   and $\chi_{\mu\nu}$ are  the Einstein tensor and massive gravity tensor  respectively \cite{cao} are given as
\begin{eqnarray}
G_{\mu\nu}&=&R_{\mu\nu}-\frac{1}{2}g_{\mu\nu}R,\\
\chi _{\mu \nu }& =&-\frac{c_{1}}{2}\left( \mathcal{U}_{1}g_{\mu \nu }-
\mathcal{K}_{\mu \nu }\right) -\frac{c_{2}}{2}\left( \mathcal{U}_{2}g_{\mu
\nu }-2\mathcal{U}_{1}\mathcal{K}_{\mu \nu }+2\mathcal{K}_{\mu \nu
}^{2}\right). 
\end{eqnarray}
The non-vanishing components of $F_{ab}$ are  $F_{\theta\phi}=g(r)\sin\theta$ and with potential $A_{\phi}=-g(r)\cos\theta$, reads \cite{AGB2}. The non-zero components of energy momentum tensor is  given by,
\begin{eqnarray}
&&T^t_t=T^r_r=\frac{8Me^{2}}{(r^2+e^2)^{\frac{5}{2}}}\\
&&T^{\theta}_{\theta}= T^{\phi}_{\phi}=\frac{8M e^2(r^2-4)}{(r^2+e^2)^{\frac{5}{2}}}.
\end{eqnarray}
Using the Eq. \ref{m2}, the $(r,r)$ components of the Eq. \ref{Field equation} are
\begin{eqnarray}
&&m'(r)+\frac{3 r^2}{2l^2}+\frac{m^2}{2}\left(\frac{cc_1 r^2}{2}+c^2c_2\right)=\frac{2Me^{2}}{{(r^2+e^2)^{\frac{5}{2}}}},
\label{rreq1}
\end{eqnarray}
where the prime is the first derivatives with respect to $r$. We can obtain
the metric function $f(r)$ using the Eq. (\ref{rreq1}). Integrating the Eq. (\ref{rreq1}) in the limit $r\to \infty$, 
\begin{equation}
m(r)+\frac{ r^3}{2l^2}+\frac{m^2}{2}\left(\frac{cc_1 r^2}{2}+c^2c_2\right)=M
\end{equation} 
 by substituting $m(r)$ in $f(r)$ Eq. (\ref{m2}), then the solution becomes
\begin{eqnarray}
f(r)=1-\frac{2M r^2}{(r^{2}+e^{2})^{\frac{3}{2}}}+\frac{ r^2}{l^2}+m^2\left (c^2c_2+\frac{cc_1 r}{2}\right) ,
\label{sol2}
\end{eqnarray}
This solutions describe the four dimensional  $AdS$ Bardeen  black hole in the presence of massive gravity and it is characterized by the mass $M$, cosmological constant $\Lambda=-3/l^2$,  magnetic charge $e$, and mass parameter $m$.   In the absence of magnetic charge $e$ this black hole solution (\ref{sol2}) interpolated with $4D$ AdS massive black hole \cite{Babichev:2015xha},
\begin{eqnarray}
f(r)=1-\frac{2M}{r}+\frac{ r^2}{l^2}+m^2\left (c^2c_2+\frac{cc_1 r}{2}\right) ,
\label{sol3}
\end{eqnarray}
 for $m=0$, it reduces to AdS Bardeen black hole as \cite{Tzikas:2018cvs}
\begin{eqnarray}
f(r)=1-\frac{2M r^2}{(r^{2}+e^{2})^{\frac{3}{2}}}+\frac{ r^2}{l^2},
\label{sol4}
\end{eqnarray}
 and for   $e=0,m=0$ it gives to AdS Schwazshild black hole.  It would be more convenient to study the horizon structure. The horizon of the black hole  can be achieved when $f(r_+)=0$. 
\begin{eqnarray}
 1-\frac{2GM r^2}{(r^{2}+e^{2})^{\frac{3}{2}}}+\frac{ r^2}{l^2}+m^2 \left(c^2c_2+\frac{cc_1 r}{2}\right)=0
\end{eqnarray}
 This equation can not be solved analytically.  The numerical analysis of the $f(r_+)=0$ on the varying the massive gravity parameters is depicted in the Fig. (\ref{fig:1}). The numerical analysis of $f (r_+ ) = 0$  reveals that it is possible to find non-vanishing
value of magnetic monopole charge ($e$),  cosmological  constant ($\Lambda$) and massive parameter ($m$)  for which metric function $f (r )$ is minimum, i.e, $f (r_+ ) = 0$ this will give four roots $r_+, r_-, r_c$ and $r_m$  which  correspond to the Cauchy horizon, event horizon, cosmological horizon and fourth horizon due to mass parameter ($m$) of the black holes respectively. The size of black hole increase with decrease the massive parameter ($m$) with constant magnetic charge ($e$). 
\begin{figure*}[ht]
\begin{tabular}{c c c c}
\includegraphics[width=.55\linewidth]{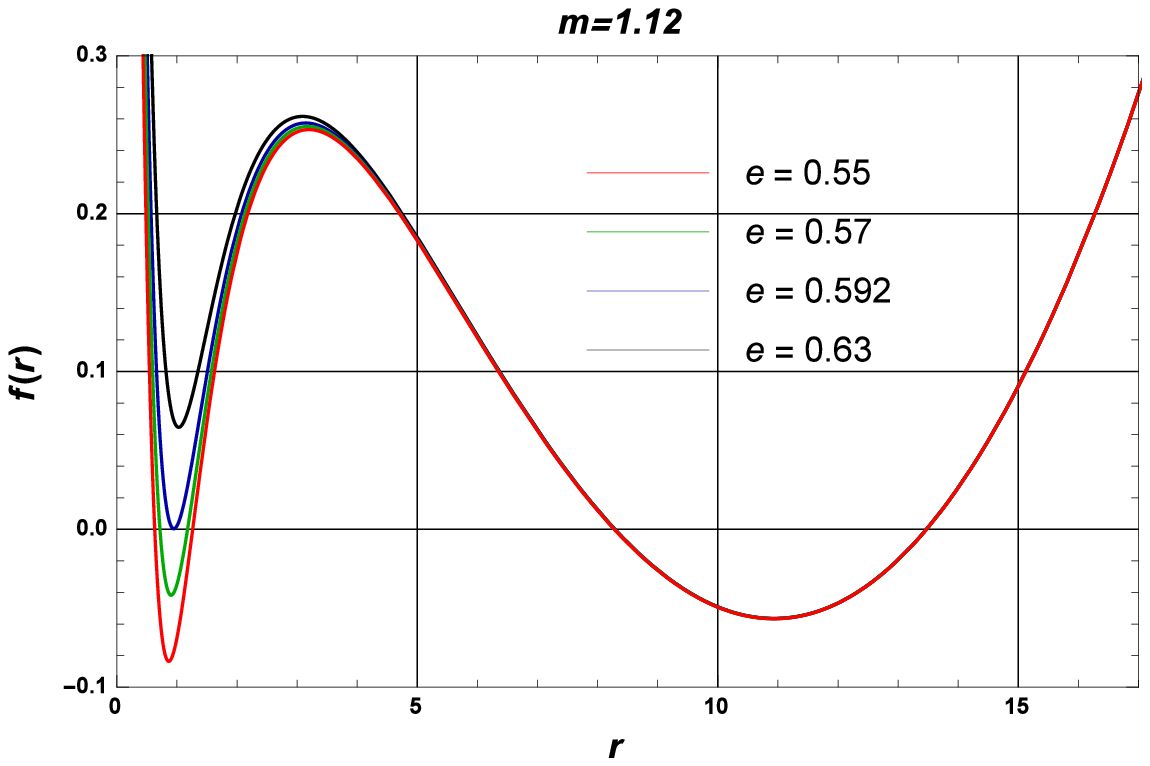}
\includegraphics[width=.55\linewidth]{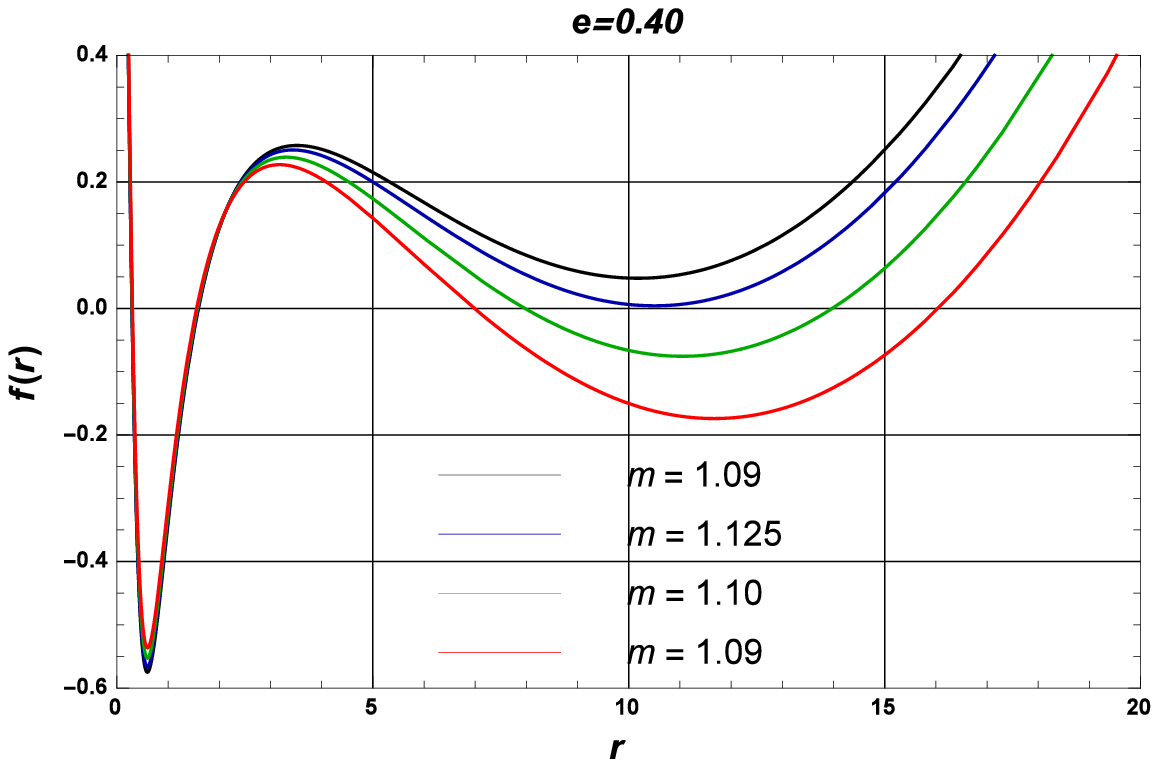}
\end{tabular}
\caption{The plot of $f(r)$ vs $r$ for different values of magnetic charge  $e$ and mass parameter $m$  with fixed value of $c=1,c_1=-1, c_2=1,l=20$ and $m=1.12$ (left) and magnetic charge $e=0.40$ (right).}
\label{fig:1}
\end{figure*}

\noindent The horizon limit of the Bardeen AdS massive black hole are: \\
1) The size of horizon increase with decrease  the magnetic charge and  the magnetic charge does not affect  the cosmological and massive horizon.\\
2)  The cosmological and massive horizon increases with the mass parameter $m$ and no cosmological and massive horizon for $m<1.125$.\\
3)  The Cauchy and event horizon independent of mass parameter $m$ and the cosmological and massive horizon independent of magnetic charge $e$.\\

Now, let us study the nature of singularity structure at $r=0$. It becomes useful to consider the curvature invariants of the spacetime such as the   Ricci square ($R_{ab}R^{ab}$), and Kretshmann scalars ($R_{abcd}R^{abcd}$)
\begin{eqnarray}
&&\lim_{r\to 0}R_{ab}R^{ab}=\frac{36}{l^2}-\frac{144M}{e^3}\left(\frac{1}{l^2}-\frac{M}{e^3}+\frac{ m^2cc_1}{4} \right)+\frac{42m^2c^2}{e^2}\left(\frac{m^2c^2c_2^2}{e^2}+\frac{2c_2}{l^2}+\frac{5c_1^2}{12}+\frac{c^4c_2^2}{e^2} \right),\nn\\
&&\lim_{r\to 0}R_{abcd}R^{abcd}=\frac{12}{l^4}-\frac{48M}{e^3}\left(\frac{1}{l^2}+\frac{M}{e^3}+\frac{5 m^2 c_2c^2}{6e^2}\right)+\frac{7m^4 c^2}{e^2}\left(c_1^2
+\frac{6 c^2c_2^2}{e^2}+\frac{2e^2c^2c_2^2}{7}+\frac{4c_2}{m^2l^2}\right).\nn\\
\label{RR}
\end{eqnarray}
These curvature invariants show that the   solution (\ref{sol2}) is regular everywhere including origin ($r=0$). The singularity of the solution is removed due to the presence of non-linear source (\ref{matter}). 

\section{\label{sec:level5} Thermodynamics}
In this section, we explore the thermodynamics of the Bardeen-AdS  massive black hole solutions (\ref{sol2}). The thermodynamical quantities temperature ($T_+$), entropy ($S_+$) and heat capacity ($C_+$) associated with the black hole solutions. The mass of the Bardeen-massive-AdS black hole  is 
\begin{eqnarray}
M_+=\frac{(r_+^{2}+e^{2})^{\frac{3}{2}}}{2r_+^2}\left[1+\frac{r_+^2 }{l^2}+{m^2}\left(c^2c_2+\frac{
cc_1r_+}{2}\right)\right]
\label{mass}
\end{eqnarray}
and the mass of   AdS Bardeen  black hole \cite{Tzikas:2018cvs} for $m \to 0$, one recovers mass of the AdS Schwarzschild black hole from (\ref{mass}), when  $m =0, e=0$. The temperature of the black hole  associated with it, known as Hawking temperature. Hawking temperature of the black hole can be defined by 
\begin{eqnarray}
T=\frac{\kappa}{2\pi}=\frac{\sqrt{-\frac{1}{2}\nabla_\mu\xi_\nu \nabla^\mu \xi^\nu}}{2\pi}=\frac{1}{4\pi}\frac{\partial{\sqrt{-g^{rr} g_{tt}}}}{\partial r}\mid_{r=r_+},
\end{eqnarray}
where $\kappa$ is surface gravity, $\xi^\mu= \partial^\mu_t$ is a Killing vector for static spherically symmetric case. Now, on inserting the temperature $T_+$ associated with the Bardeen massive black hole can  be calculated as
 \begin{eqnarray}
T_+ =\frac{1}{4\pi r_+( e^2+r_+^{2})}\left[{r_+^2- e^2+\frac{m^2}{2}(c_2c^2r_+(2r^2_+-e^2)+2r_+c c_1(r_+^2-2e^2))}+\frac{3r_+^2}{l^2}\right],
\label{temp1}
\end{eqnarray}
and the  AdS Bardeen  black hole \cite{Tzikas:2018cvs}, when $m \to 0$, one recovers from (\ref{temp1}) the temperature of   Bardeen black hole \cite{dvs19}
\begin{equation}
T_+= \frac{1}{4\pi r_+}\left(\frac{r_+^2-e^2}{r_+^2+e^2}\right).
\label{eqT1}
\end{equation}
In turns the temperature reduces to $T_+=1/4\pi r_+$ of the  Schwarzschild  black holes, can be obtained by taking   $e=0$ in (\ref{eqT1}). The Fig. \ref{fig:1} displays the behaviour of temperature for various values of $e$ and $m$, which  shows that the temperature grows to a maximum $T_{max}$ then it drop to minimum value and increase again. The maximum temperature $T_{max}$ depends on charge $e$ and is shown in the Table \ref{tab:temp}.  Thus a remarkable result is that, unlike the  Schwarzschild black hole, for the  radius corresponding to the  temperature of Bardeen AdS massive black hole  increases with $e$ and decreses with the mass parameter $m$(See in the Fig. \ref{fig1}).  A maximum of  Hawking temperature occurs at the critical radius shown in Table \ref{tab:temp}. It turns out that the maximum temperature decreases with increase in the values of $e$ and $m$ which diverges when the horizon radius shrinks to zero.
\begin{figure*}[ht]
\begin{tabular}{c c c c}
\includegraphics[width=.55\linewidth]{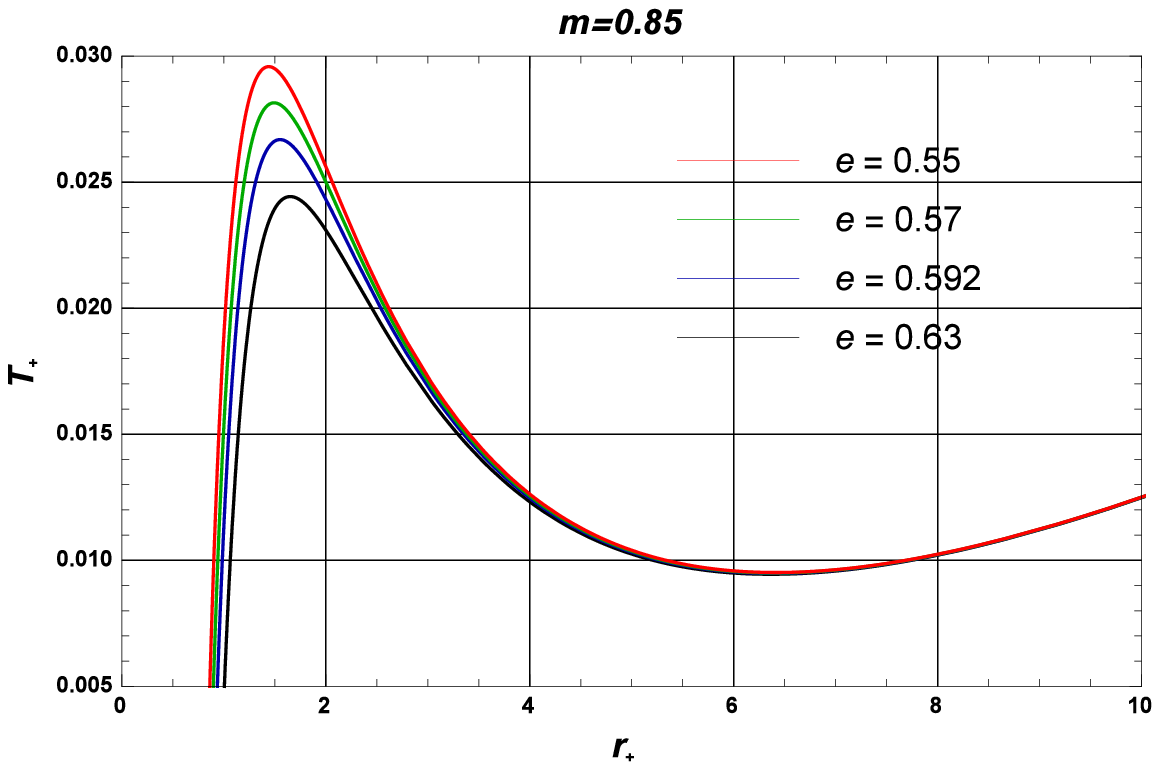}
\includegraphics[width=.55\linewidth]{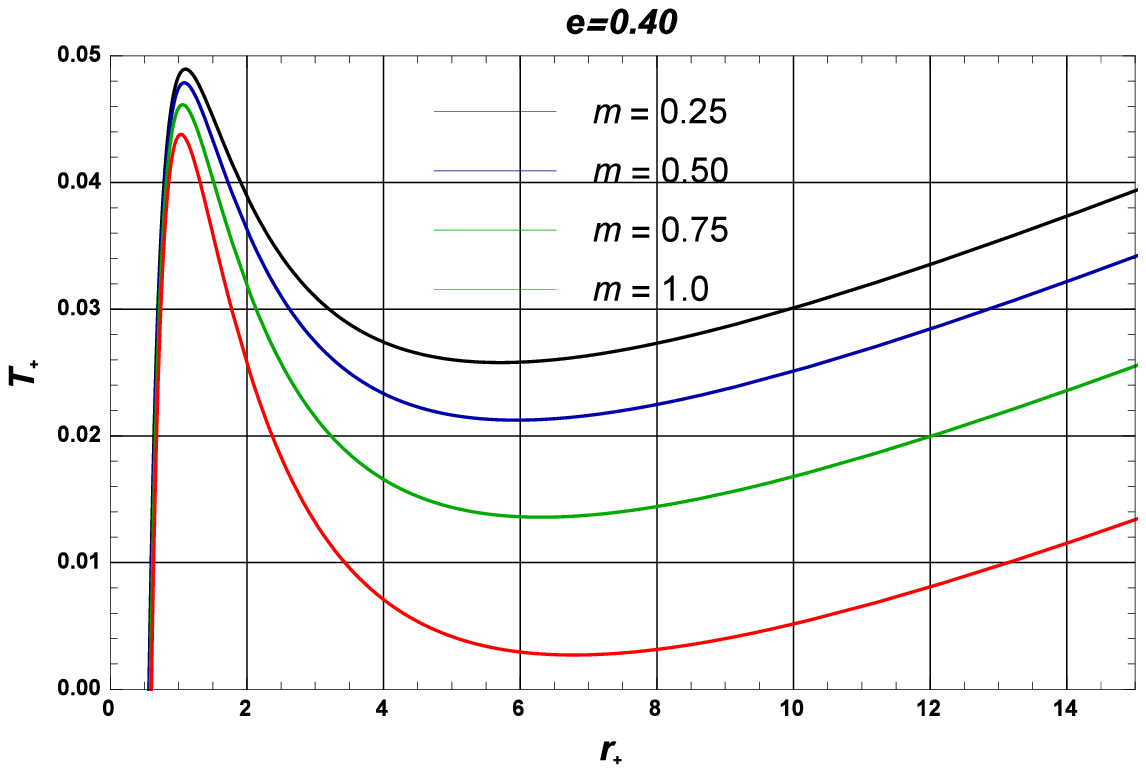}\\
\end{tabular}
\caption{The plot of temperature vs horizon radius $r_+$ with different value of magnetic charge $e$ (Left) and different value of mass parameter $m$ (Right) with fixed value of $c=1,c_1=-1, c_2=1$ and $l=20$. }
\label{fig:1}
\end{figure*}

\begin{center}
	\begin{table}[h]
		\begin{center}
			\begin{tabular}{l|l r l r l| r l r r r}
				\hline
				\hline
				\multicolumn{1}{c}{ }&\multicolumn{1}{c}{ }&\multicolumn{1}{c}{ }&\multicolumn{1}{c}{m= fixed  }&\multicolumn{1}{c}{ \,\,\,\,\,\, }&\multicolumn{1}{c|}{ }&\multicolumn{1}{c}{  }&\multicolumn{1}{c}{ }&\multicolumn{1}{c}{e=fixed}\,\,\,\,\,\,\\
				\hline
				\multicolumn{1}{c|}{ } &\multicolumn{1}{c}{e= 0.55 } &\multicolumn{1}{c}{ 0.57 } & \multicolumn{1}{c}{ 0.592 }& \multicolumn{1}{c}{0.63}& \multicolumn{1}{c|}{} &\multicolumn{1}{c}{m= 0.25}&\multicolumn{1}{c}{0.50} &\multicolumn{1}{c}{ 0.75}   & \multicolumn{1}{c}{1.0}& \multicolumn{1}{c}{} \\
				\hline
				$r_c^T$ & 1.485& 1.531 &1.1578 &1.647 &&1.12\,\,\,&1.11&1.05&1.03&
				\\
				$T_+^{Max}$&0.0296& 0.0281 &0.0266& 0.0243 &&0.0255& 0.0248&0.0235&0.0228&\\
				\hline
				\hline
			\end{tabular}
		\end{center}
		\caption{Maximum Hawking temperature ($T_+^{max}$) at critical radius ($r_c^{T}$) for the $4D$ AdS Bardeen massive black hole.}
		\label{tab:temp}
	\end{table}
\end{center}
\noindent One can easily obtain the following formula by using the first
law of thermodynamics $dM_+=T_+dS_++\phi de$ for the entropy of black hole
\begin{eqnarray}\label{sf}
S_+=\int T_+^{-1} dM_+=\int T_+^{-1}\frac{d M_+}{d r_+}dr_+.
\end{eqnarray}
Now, substituting the value of $M_+$ and $T_+$  form Eqs. (\ref{mass}), (\ref{temp1}) into Eq. (\ref{sf}), we obtained the entropy of Bardeen AdS -massive black hole as
\begin{eqnarray}
S_+&&=\pi\Big[(r_+^2-2e^2)(r_+^2+e^2)^{1/2}+3e^2r_+\text{log}[r_++(r_+^2+e^2)^{1/2}]\Big].
\label{entropy}
\end{eqnarray}
Thus, the area law is no longer valid for regular black holes.   We can also be derived the temperature according to the entropy using the first law of thermodynamics.

The  thermodynamic or local stability of a black hole is performed by studying the behaviour of  its heat capacity ($C_+$). The  positive ($C_+>0$) heat capacity indicates that the  black hole is stable; when it is negative ($C_+<0$), the black hole is said to be unstable \cite{dvs19,kumar19,sabir}.    The heat capacity of the black hole is defined as
\begin{eqnarray}\label{SH}
C_+&=&\frac{\partial{M_+}}{\partial{T_+}}=\left(\frac{\partial{M_+}}{\partial{r_+}}\right)\left(\frac{\partial{r_+}}{\partial{T_+}}\right).
\end{eqnarray}
 The heat capacity of Bardeen -AdS-massive  black hole, by using Eqs.  (\ref{mass}), (\ref{temp1}), and (\ref{SH}),  we obtained
\begin{widetext}
\begin{eqnarray}
C_+&=&\frac{\pi (e^2+r_+)^{5/2}(6 r_+^4+2l^2(2r_+^2-e^2)+m^2l^2(2c^2c_2r_+^2+2cc_1r_+^3-4c^2c_2e^2-cc_1e^2)}{r_+(2e^4l^2+7e^2l^2r_+^2+9e^2r_+^4-l^2r_+^4+3r_+^6)+m^2l^2(2c^2c_2e^4+7c^c_2e^2r_+^2+cc_1e^2r_+^3-c^2c_2r_+^4)},\nonumber\\
\label{SH1}
\end{eqnarray}
\end{widetext}

\begin{figure}[ht]
\begin{tabular}{c c c c}
\includegraphics[width=.5\linewidth]{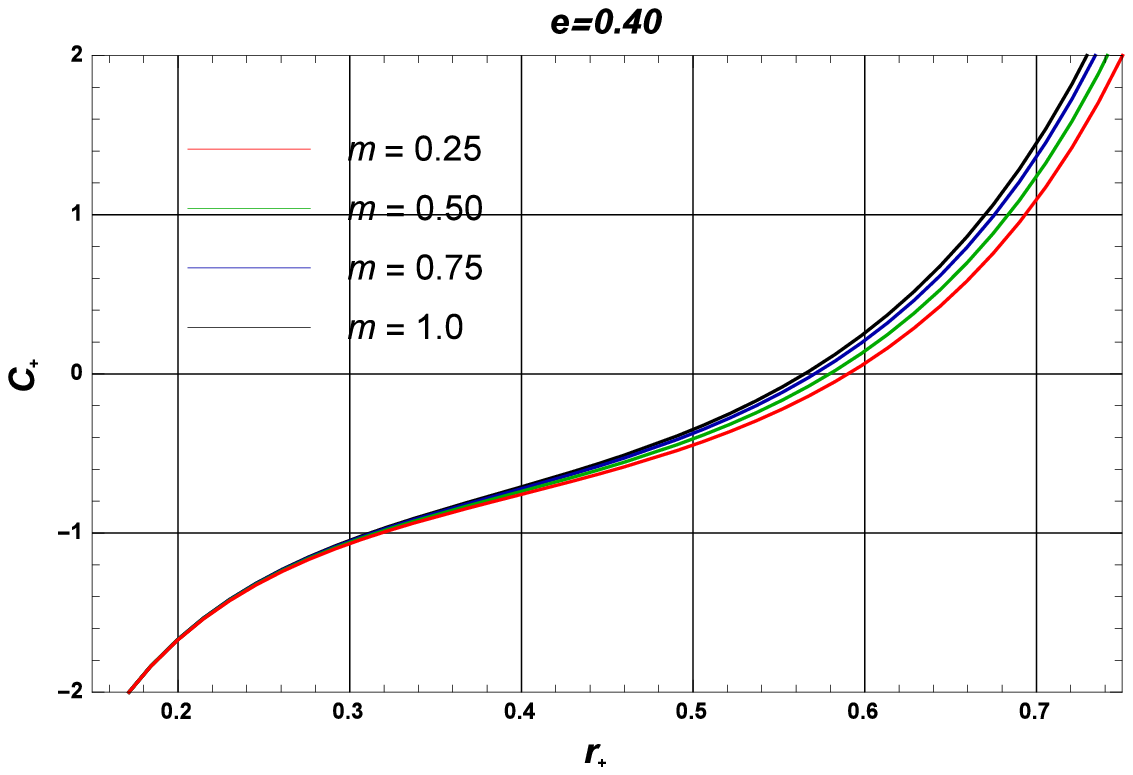}
\includegraphics[width=.5\linewidth]{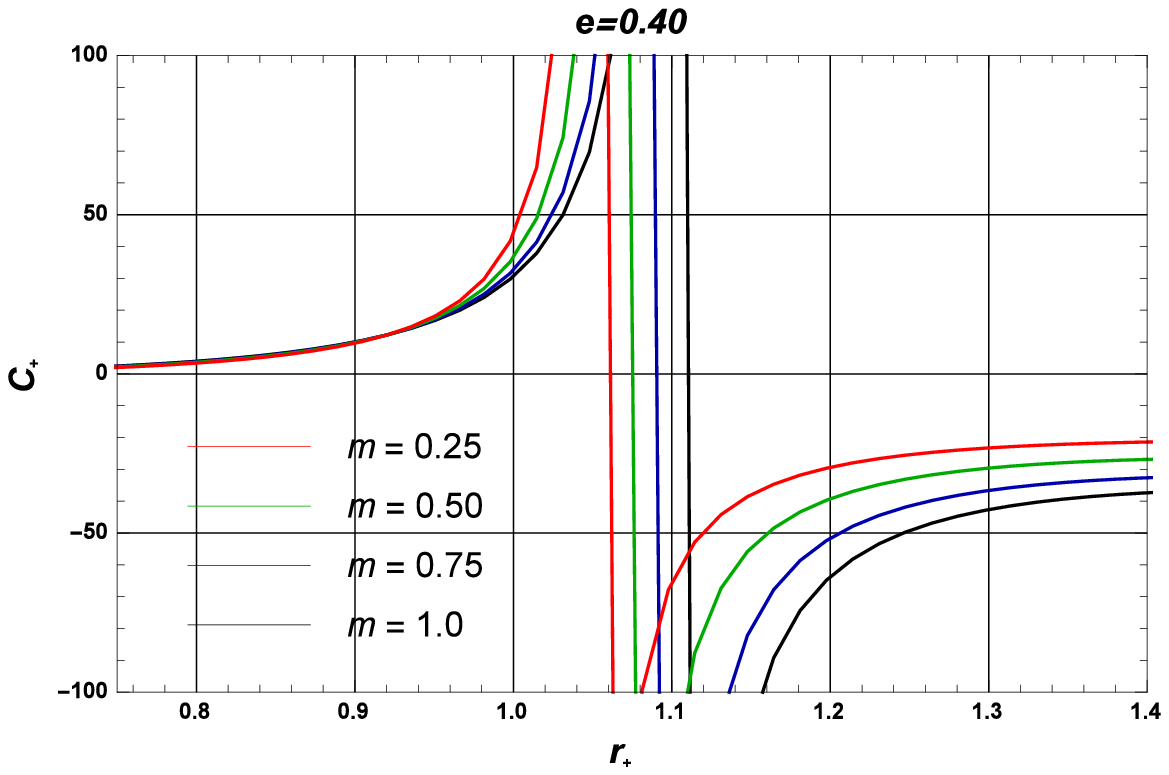}\\
\includegraphics[width=.5\linewidth]{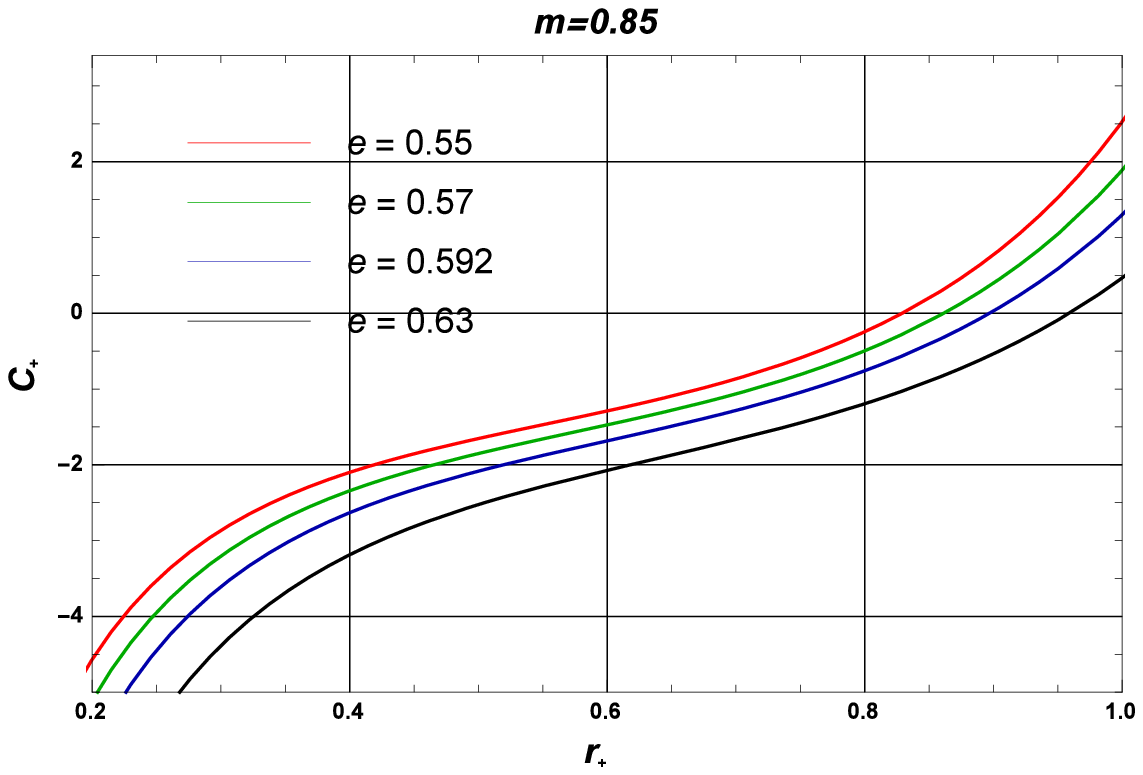}
\includegraphics[width=.5\linewidth]{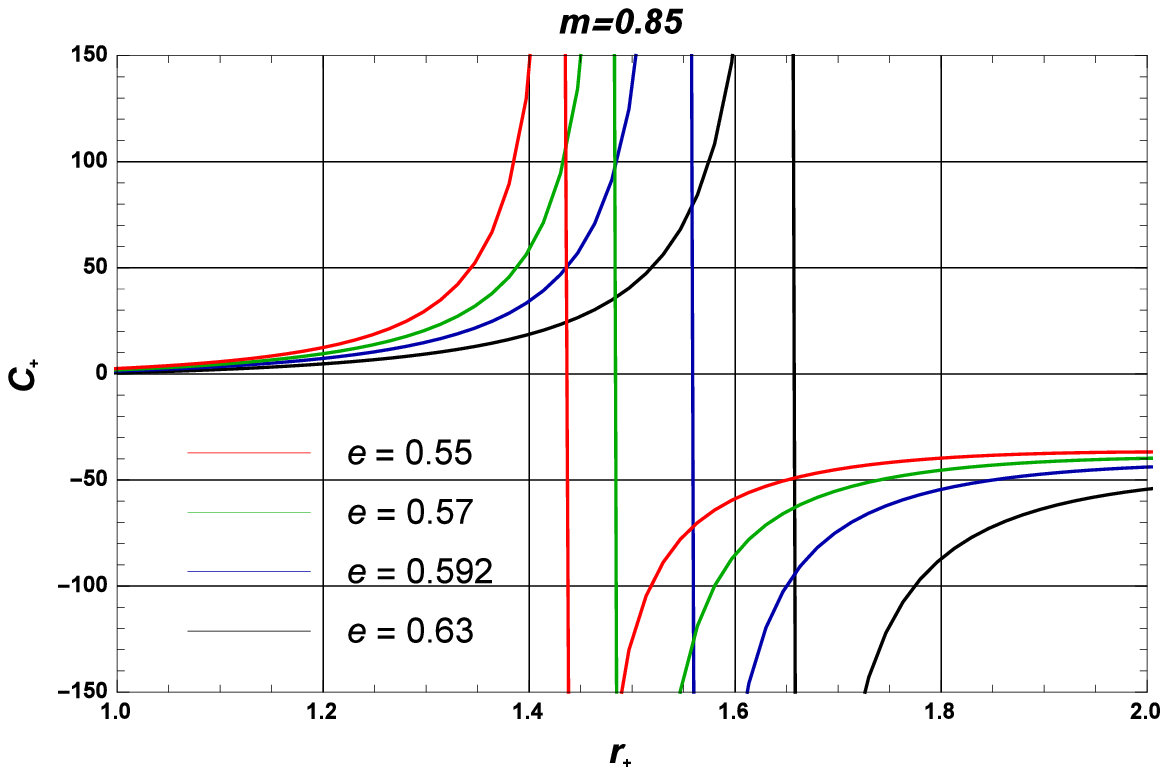}
\end{tabular}
\caption{The plot of specific heat vs horizon radius  with fixed value of $c=1,c_1=-1, c_2=1, e=0.40$ and $l=20$. }
\label{fig:1}
\end{figure}

It can be seen clearly, that the heat capacity depends on the charge ${e}$,  mass parameter $m$, and  cosmological constant $\Lambda$.   The heat capacity is plotted  in Fig. \ref{fig:1} for different values of $e$. The  heat capacity is positive (negative) $r_+<r_C$ $ (r_+>r_C) $ suggesting  thermodynamic stability (instability) of smaller (larger) regular black holes. The heat capacity is discontinuous  at $r_+=r_C$ which means the second order phase transition occurs  \cite{hp,dvs99}. Interestingly, the discontinuity of the heat capacity occurs at $r_+=1.439$, point at which point  the Hawking temperature has the  maximum value $T_+=0.0296$ for  $e=0.55$. Hence the phase transition occurs from the lower to higher mass black holes corresponding from positive to negative heat capacity of the black hole.  In the absence of  mass parameter $m$, it reduce to the expression for heat capacity of  Bardeen  black hole  \cite{dvs19}.    The  heat capacity for Schwarzschild  black hole $C_{+} =-2\pi r^{3}_+.$  in the limit $m=0$ .

\section{\label{sec:level6} Thermodynamic Stability and Phase Diagrams}
\noindent Now we study the phase transition of the Bardeen AdS massive black hole by using the P-V criticality and phase diagrams.  The thermodynamical relation between the pressure and cosmological constant given as
\begin{equation}
P=-\frac{\Lambda}{8\pi}=\frac{3}{8\pi l^2}.
\label{lpre}
\end{equation}
  The mass of  the black hole interpretes  enthalpy of thermodynamical system. The equation of state can by obtained by using the Eq. (\ref{temp1})  and  Eq. (\ref{lpre}) as
\begin{eqnarray}
P_+=\frac{1}{2r_+ T_+}\left(1+\frac{e^2}{r_+^2}\right)-\frac{1}{4\pi r_+^2}\left(1-\frac{e^2}{2r_+^2}\right)+\frac{m^2}{16\pi r_+^4}\left(cc_1r_+(e^2-2r_+^2)+2c^2c_2(r_+^2-2e^2)\right),\nonumber\\
\label{pv4}
\end{eqnarray}
and the corresponding thermodynamics volume is
\begin{eqnarray}
 V=\left(\frac{\partial H_+}{\partial P_+}\right)_{S_+,e}= \frac{4\pi}{3}{\left(r_+^{2}+e^{2}\right)^{\frac{1}{2}}}
\label{pv3}
\end{eqnarray}
At the  inflection point, we can determine the critical temperature and critical  pressure b following equations 
\begin{equation}
\frac{\partial P_+}{\partial r_+}=0,\qquad \qquad \frac{\partial ^2P_+}{\partial r_+^2}=0.
\label{pv5}
\end{equation}
Substitution Eq.  (\ref{pv4}) into Eq.  (\ref{pv5}), we find the critical points and  the  horizon radius satisfy the following equation
\begin{eqnarray}
\frac{24e^2+30e^2r_+^2-2r_+^4+m^2[c_2c^2(24e^4+30r_+^2e^2-2r_+^4)+9cc_1e^2r_+^3]}{128\pi^3r_+^{8}(e^2+r_+^2)(3e^2+r_+^2)}.
\label{pv6}
\end{eqnarray}
The  Eq. (\ref{pv6}) can not be solved analytically, so we can calculate the critical radius $r_+$, critical pressure $P_+$ and temperature  $T_+$  numerically and the numerical results are presented in  Table \ref{tr10} and  Table \ref{tr12} for different value of mass parameter $m$ and magnetic charge $e$.  We can see that the  critical  radius $r_+$ increases with magnetic charge $e$   and decrease with the mass parameter $m$. The universal ratio $P_cr_c/T_c$ are increasing function of the magnetic charge $e$ and  massive parameter $m$. 
\begin{table}[ht]
 \begin{center}
 \begin{tabular}{ l | l   | l   | l   | l   }
\hline
            \hline
  \multicolumn{1}{c|}{ $m$} &\multicolumn{1}{c}{$r_C$}  &\multicolumn{1}{|c|}{$T_C$}  &\multicolumn{1}{c|}{$P_C$} &\multicolumn{1}{c}{$\frac{P_C\,r_C}{T_C}$}\\
            \hline
            \,\,\,\,\,0 ~~  &~~0.3970~~  & ~~0.2513~~ & ~~0.1161~ & ~~0.1821~~ \\    
            \,\,\,\,\,1~~ &~~0.3864~~ & ~~0.4425~~ & ~~0.2489~~ &  ~~0.2173~~      \\
            \,\,\,\,\,2~~ &~~0.3803~~  & ~~1.0174~~ & ~~0.6488~~ &  ~~0.2425~~    \\
            \,\,\,\,\,3~~ &~~0.3782~~ & ~~1.9758~~ & ~~1.3158~~ &    ~~0.2518~~  \\
            \,\,\,\,\,4~~ &~~0.3774~~  & ~~3.3178~~ & ~~2.2496~~ &    ~~0.2558~~   \\
            \,\,\,\,\,5~~ &~~0.3770~~  & ~~5.0432~~ & ~~3.4504~~ &    ~~0.2579~~    \\
            \hline 
\hline
        \end{tabular}
        \caption{The table for critical temperature $T_C$, critical pressure $P_C$ and $P_C\,r_C/T_C$ corresponding different value of mass $m$ with fixed value of $c=1,c_1=-1, c_2=1,e=0.1$ .}
\label{tr10}
    \end{center}
\end{table}
\begin{table}
 \begin{center}
 \begin{tabular}{ l | l   | l   | l   |  l   }
\hline
            \hline
  \multicolumn{1}{c|}{ $e$} &\multicolumn{1}{c|}{$r_C$}  &\multicolumn{1}{c|}{$T_C$}  &\multicolumn{1}{c|}{$P_C$} &\multicolumn{1}{|c}{$P_C\,r_C/T_C$}\\
            \hline
            \,\,\,\,\,0.1 ~~  &~~0.3864~~  & ~~0.4425~~ & ~~0.2489~~ & ~~0.1821~~ \\            
            \,\,\,\,\,0.2~~ &~~0.7524~~ & ~~0.1917~~ & ~~0.0667~~ &  ~~0.2617~~      \\
            \,\,\,\,\,0.3~~ &~~1.099~~  & ~~0.1083~~ & ~~0.0317~~ &  ~~0.3216~~    \\
            \,\,\,\,\,0.4~~ &~~1.427~~ & ~~0.06697~~ & ~~0.0191~~ &    ~~0.4069~  \\
            \,\,\,\,\,0.5~~ &~~1.738~~  & ~~0.04231~~ & ~~0.0131~~ &    ~~0.5381~~   \\
            \hline 
\hline
        \end{tabular}
        \caption{The table for critical temperature $T_C$, critical pressure $P_c$ and $P_C\,r_C/T_c$ corresponding to the different value of  magnetic charge $e$ with fixed value of $c=1,c_1=-1, c_2=1, m=1$ .}
\label{tr12}
    \end{center}
\end{table}
In order to obtained the phase transition of the black holes an analog with the vander Walls phase transition, we can identify the free energy of the black hole. to see the effect of magnetic charge $e$ and mass parameter $m$ on the phase structure of the system, we fix the magnetic charge $e$ first and vary the mass parameter $m$. The   critical pressure and  critical temperature  increase  with critical radius (see Fig. \ref{tr10}) and the opposite nature show when we fix the mass parameter $m$ and vary of  magnetic charge $e$.

\begin{figure}[ht]
\begin{tabular}{c c c c}
\includegraphics[width=.5\linewidth]{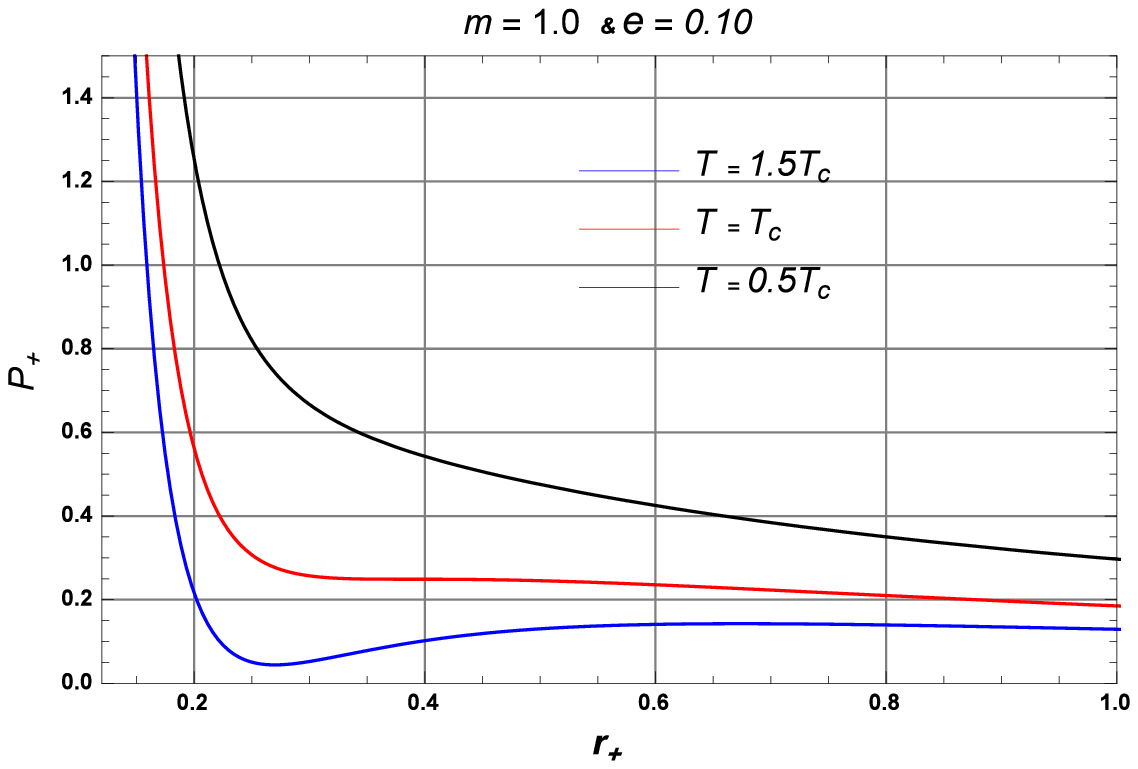}
\includegraphics[width=.5\linewidth]{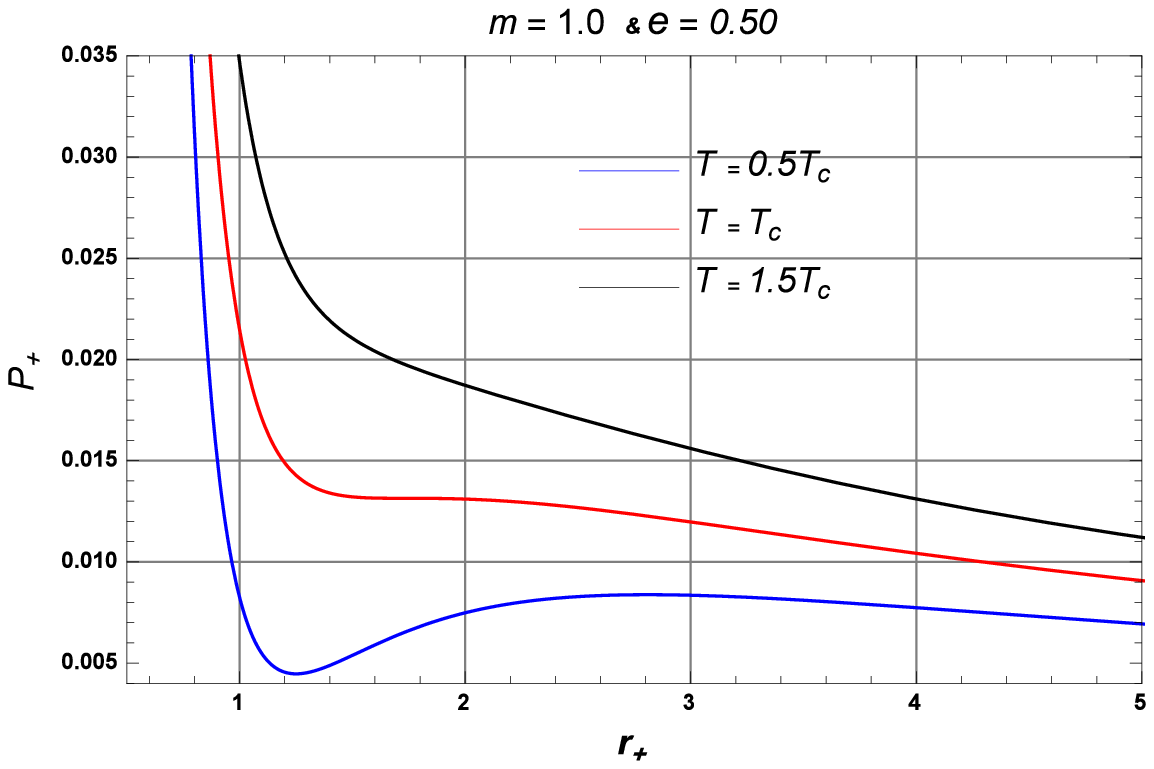}\\
\includegraphics[width=.5\linewidth]{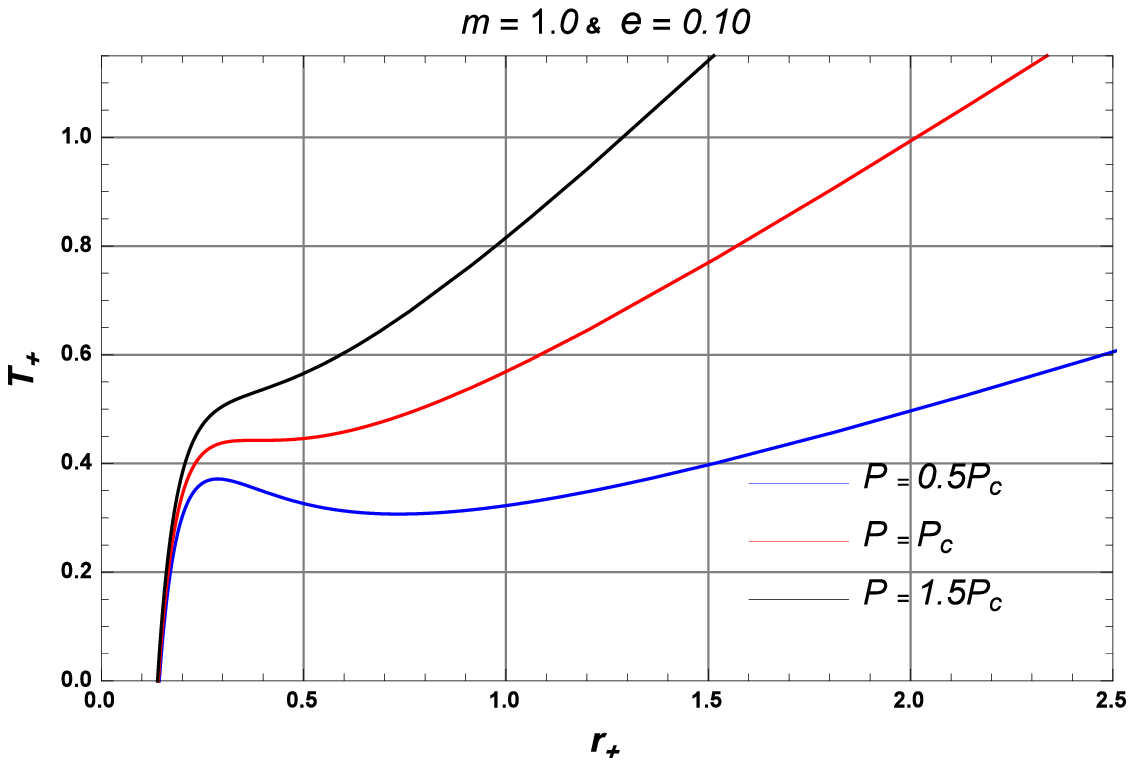}
\includegraphics[width=.5\linewidth]{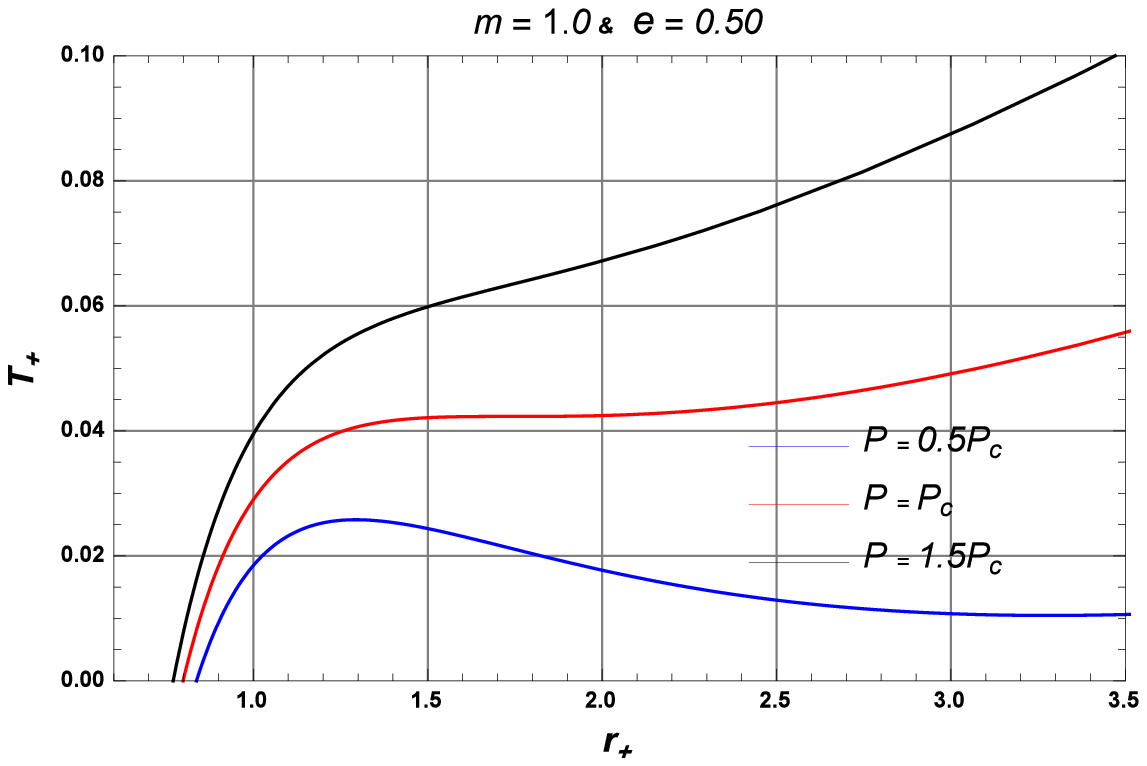}\\
\includegraphics[width=.49\linewidth]{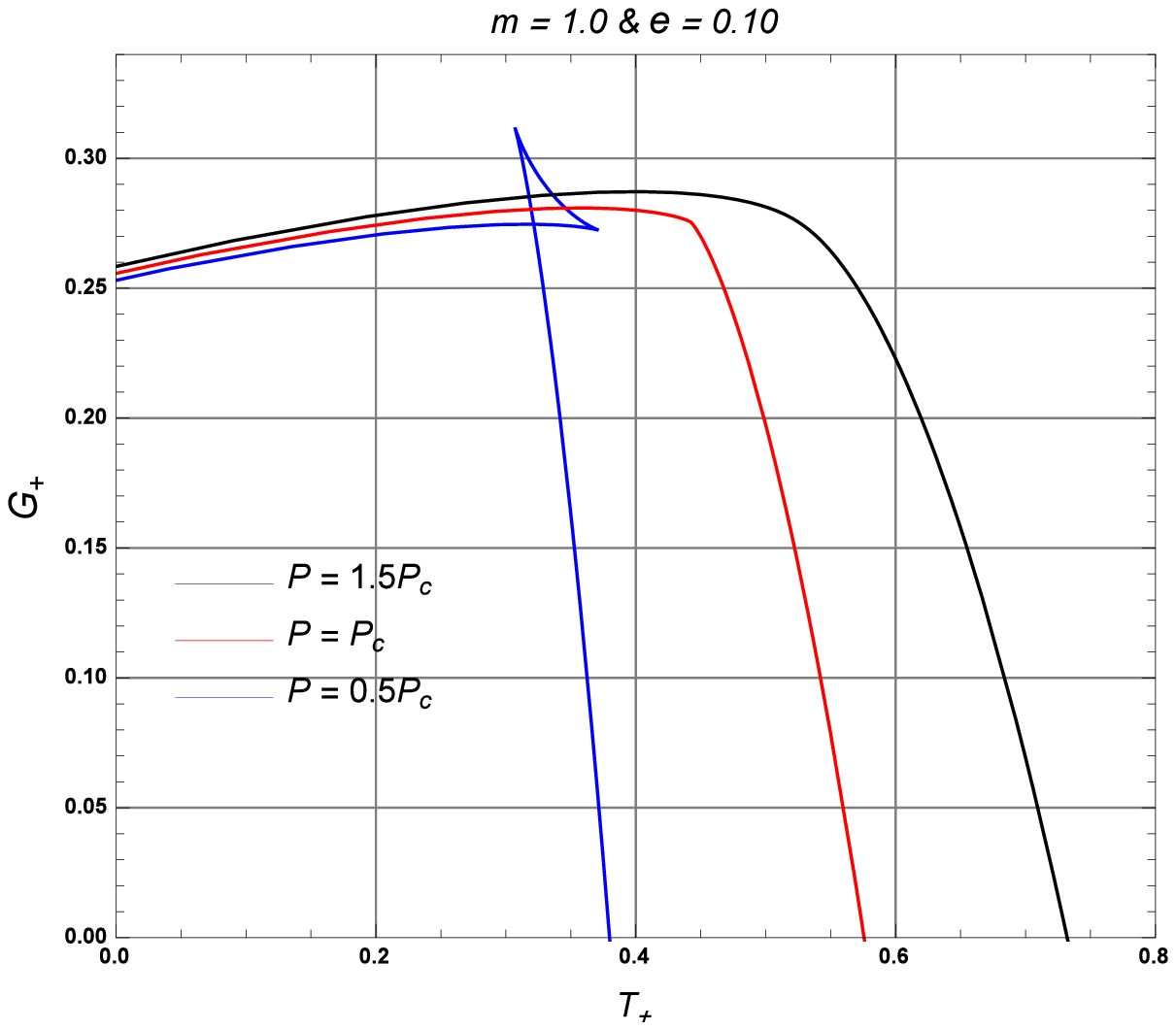}
\includegraphics[width=.49\linewidth]{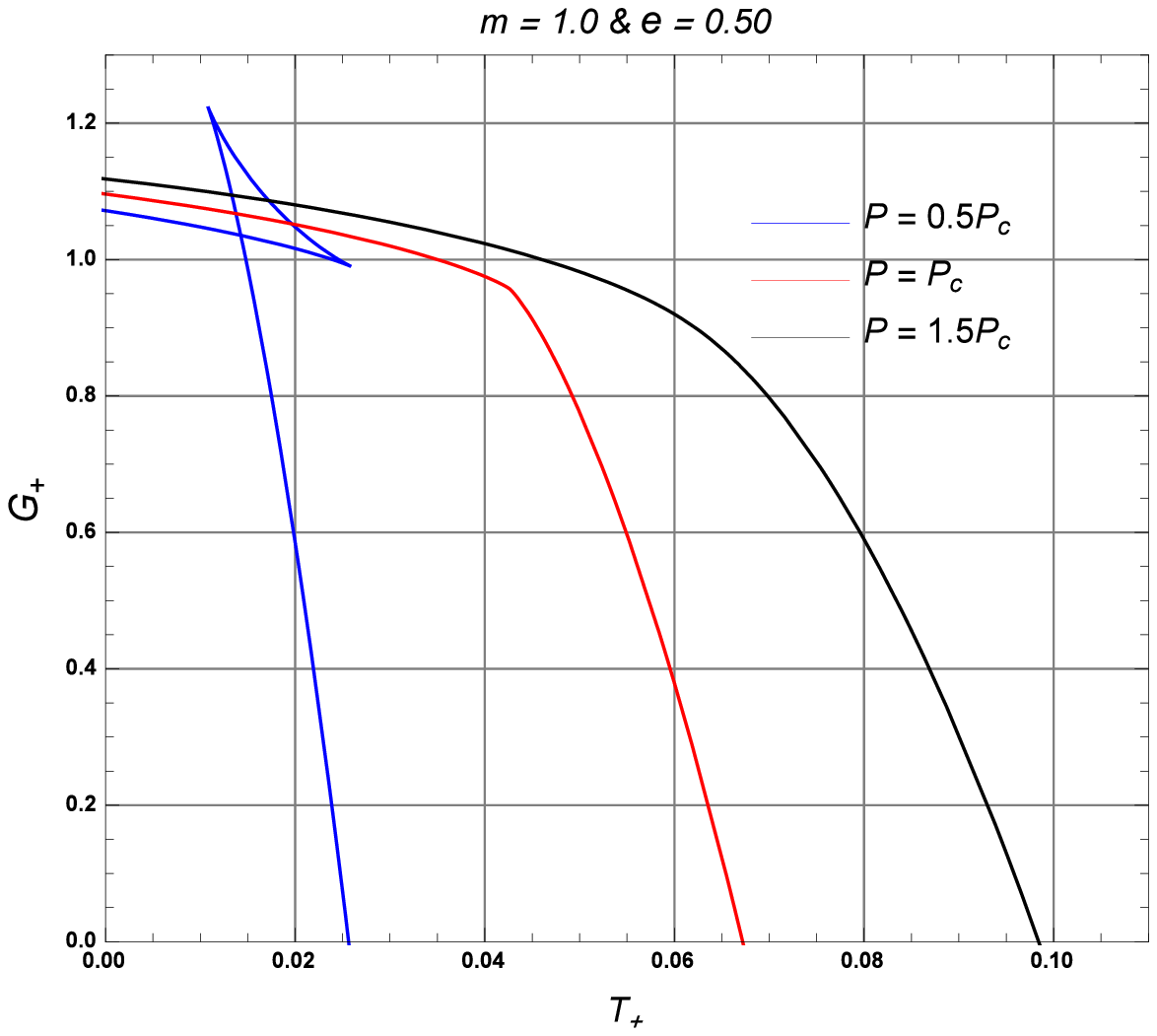}
\end{tabular}
\caption{The plots of pressure and temperature vs horizon radius  and Gibbs free energy ($G_+$) vs temperature ($T_+$)  for $e=0.1$ (left) and $e=0.5$ (right) with fixed values of $c=1,c_1=-0.75,c_2=0.75$ and $m=1$ .}
\label{fig1}
\end{figure}

 In $ T_+-r_+$  plots  we can  see that the three  black hole (small, intermediate and large )  for the $P< P_c$, $P=P_c$ and $P> Pc$ for  Bardeen AdS   massive black hole with  the same magnetic charge $e$ and the mass parameter $m$ for a certain range of temperature. The small and large are stable but intermediate black hole is unstable, since the heat capacity is negative (see  Fig. \ref{fig:1}). When $ T_+<T_{\star} $ the small black hole and $ T_+>T_{\star} $  corresponding to large black hole due to small free energy.  We can transit one phase to other phase at critical temperature due to same free energy.  In    Fig. \ref{fig1} (middle)  isotherms represents the first order $T_+<T_c$ and second order  phase transition $T_+=T_{\star}$, which is obtained from the free energy diagram (see Fig. \ref{fig1} (lower)  and  the corresponding temperature is $T_{\star}=0.335$.  In $G_+ - T_+$  plots the appearance of characteristic swallow tail in   show that the obtained values are critical ones in which the phase transition take place .  In  Fig. \ref{fig1} (lower) , we can see that swallow tail shape $P<P_c$,  for the first order phase transition and $P=P_c=0.0249$ for the second order phase transition.

\begin{figure}[ht]
\begin{tabular}{c c c c}
\includegraphics[width=.5\linewidth]{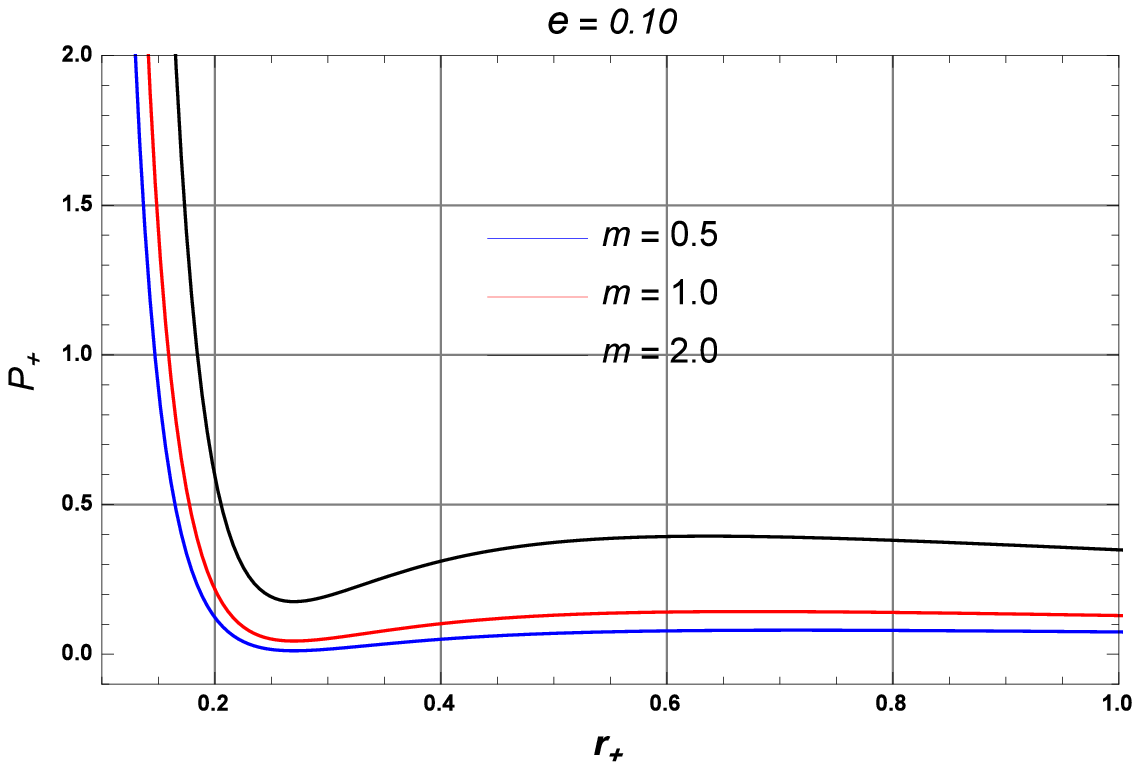}
\includegraphics[width=.5\linewidth]{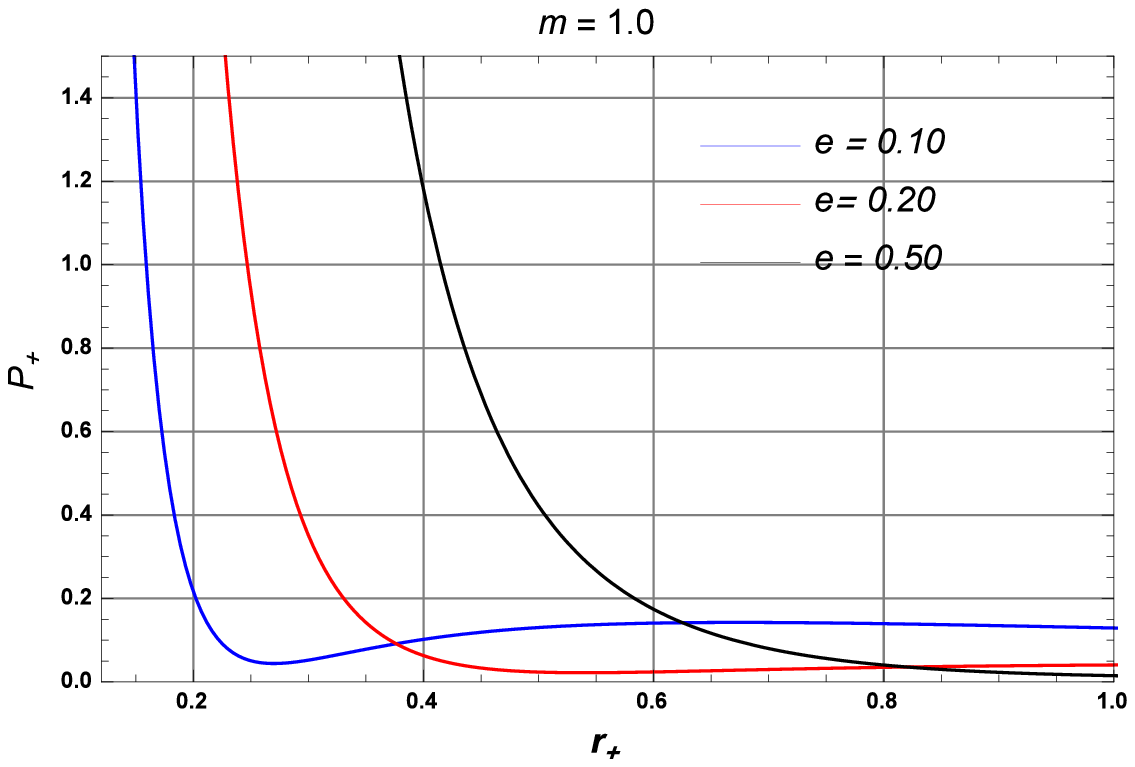}\\
\includegraphics[width=.5\linewidth]{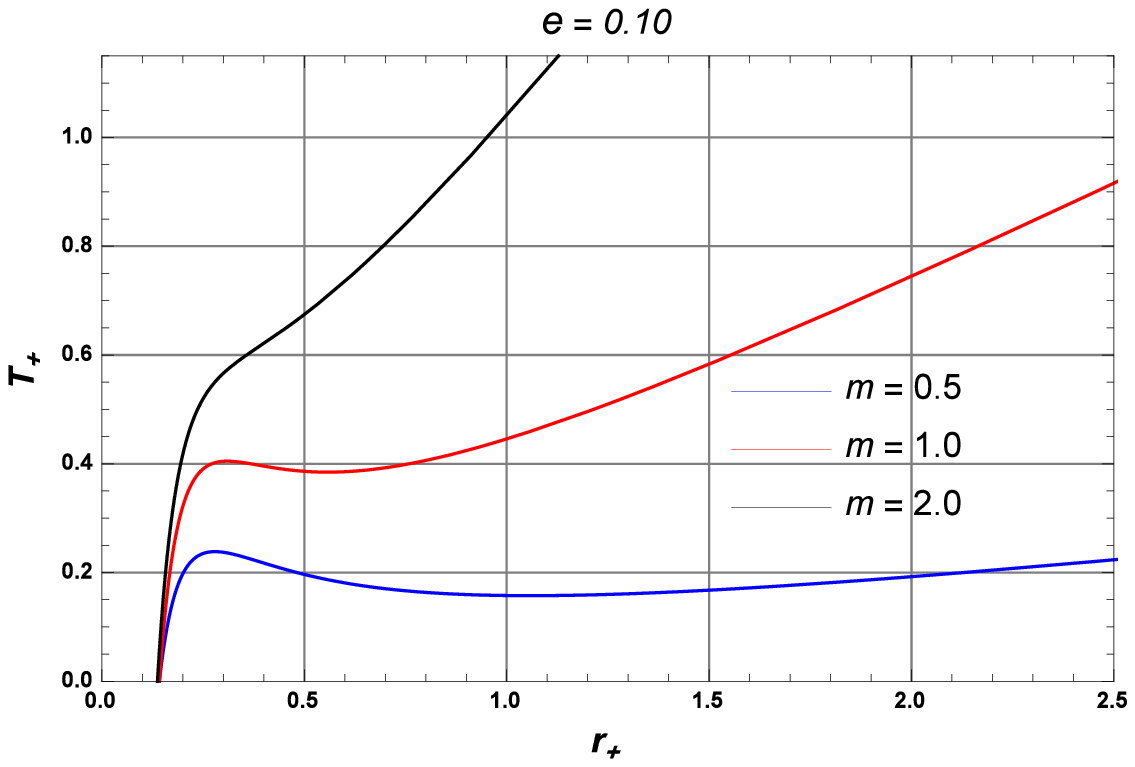}
\includegraphics[width=.5\linewidth]{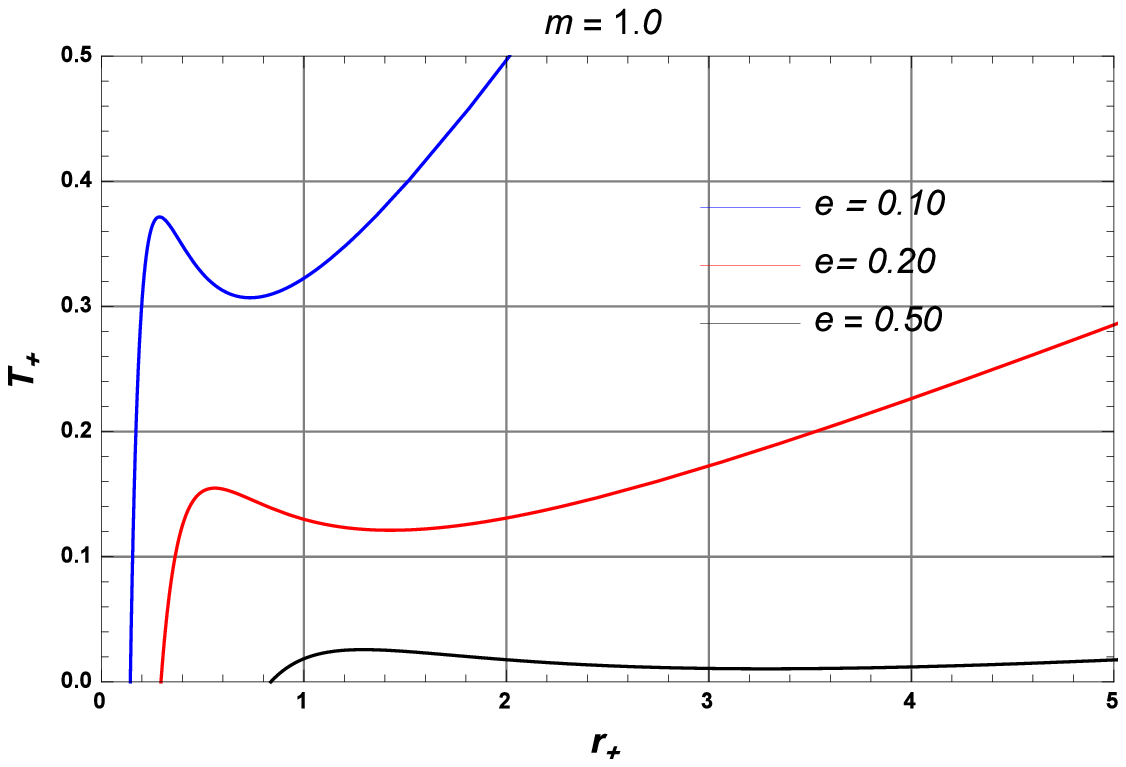}\\
\includegraphics[width=.5\linewidth]{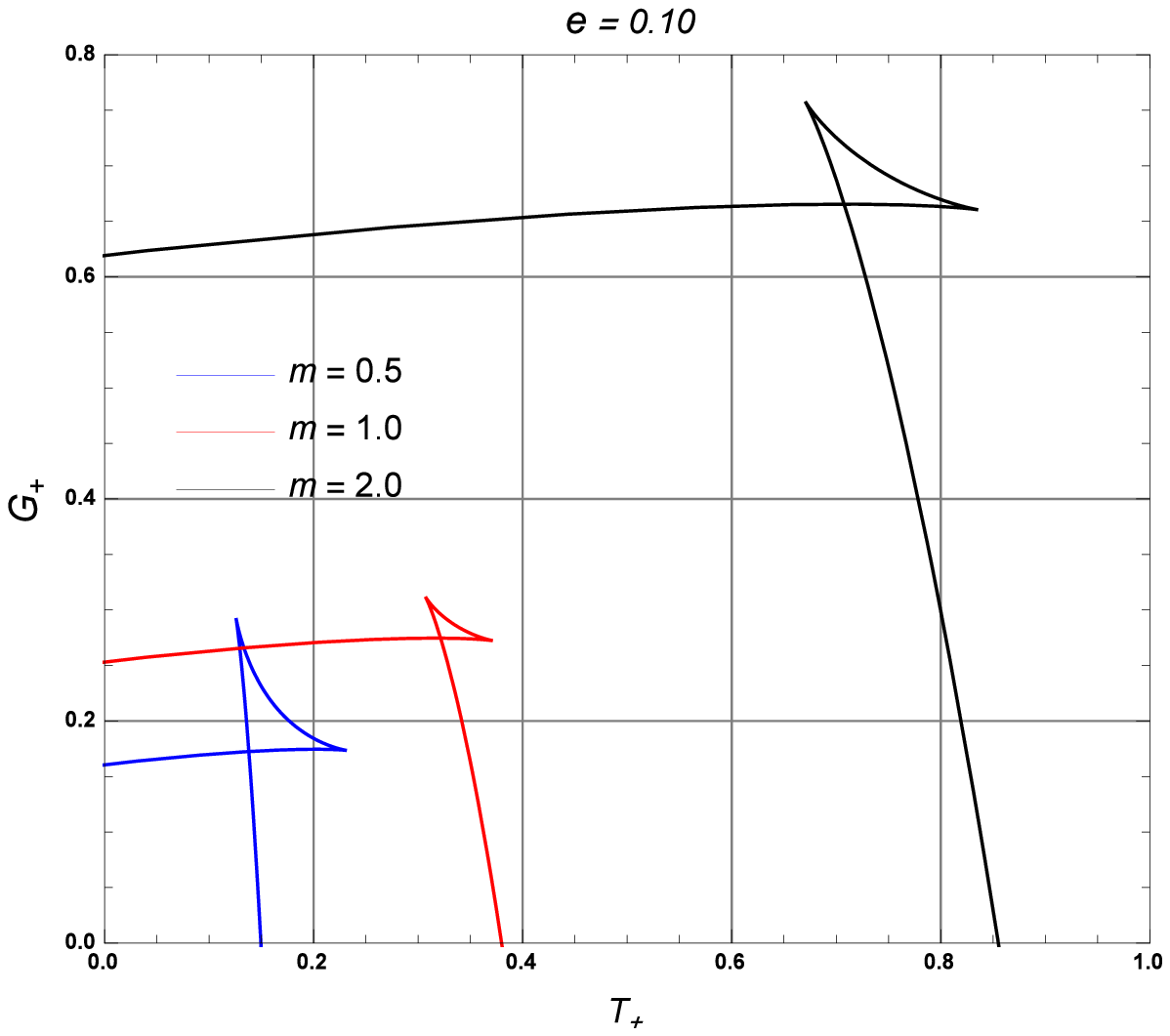}
\includegraphics[width=.5\linewidth]{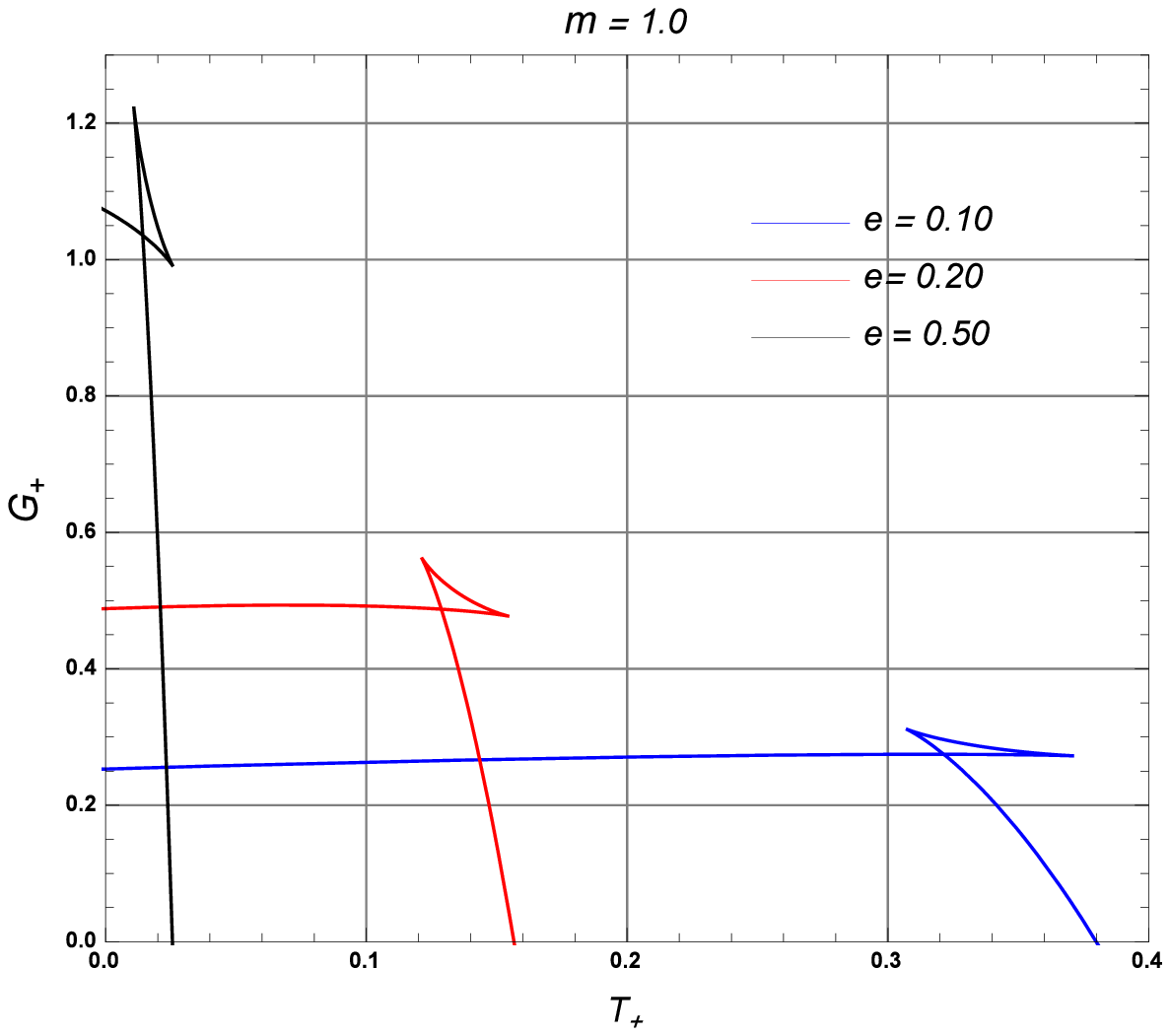}
\end{tabular}
\caption{The plots of pressure and temperature vs horizon radius  and Gibbs free energy ($G_+$) vs temperature ($T_+$) for $e=0.1$ (left) and $e=0.5$ (right) with fixed values of $c=1,c_1=-0.75,c_2=0.75$ and $m=1$ . }
\label{fig2}
\end{figure}

Fig. 5 shows   the effects of variation of mass parameter $m$ and magnetic charge $e$ on critical values.  The isobars are increase function of mass parameter $m$ and  decreasing with the magnetic charge $e$  respectively. The gap between the two sub-critical bar are the increases      
 mass parameter $m$ as well as  magnetic charge $e$. The sub-critical isobar is the region, where the phase transition take place. The only difference is that Gibbs free energy of two phases are increasing function of massive parameter $m$ and decreasing function of magnetic charge $e$.  Interestingly, the universal relation $P_c\,r_c/T_c$, shows that the the effects of variation of mass parameter $m$ and magnetic charge $e$ are decreasing function massive parameter whereas critical temperature   and pressure are increasing function of it.

\section{\label{sec:level6} Conclusions}
In this paper, we have presented exact  solution of Bardeen AdS  black holes in massive gravity with a negative cosmological constant, thereby generalizing bardeen black holes which are included as a special case ($e=0$, $l^2 \to \infty$). The AdS Bardeen black holes are characterized by analyzing horizons, which at most could be four, viz. inner Cauchy , event, cosmological  and massive horizons. We have analysed the thermodynamical properties and phase structure of  Bardeen AdS massive  black holes.  Despite complicated solutions,  exact expression for the thermodynamical quantities like the black hole mass, Hawking temperature, entropy and free energy at event horizon $r_+$ are obtained. The thermodynamical stability of the black holes is also analysed by studying the  heat capacity. The entropy (\ref{entropy}) of the black holes is modified due to the magnetic charge $e$ and the mass parameter $m$ resulting area law $S=A/4$ no longer valid. The phase transition is detectable by the divergence of the heat capacity ($C_+$) at a critical radius $r_c$ (changes with $e, m$ ), such that the black hole is stable  with positive heat capacity ($C_+>0$), and unstable with negative heat capacity ($C_+<0$).

The phase transition of  Bardeen AdS massive black holes have also been studied in the extended phase by considering the cosmological constant as thermodynamic pressure.  It is seen  that the thermodynamic volume is independent of  mass parameter $m$ and magnetic charge $e$. The phase diagram shows that the obtained critical values of pressure and temperature are one in which phase transition take place.   Interestingly, it is pointed out that the nature of  massive parameter $m$ and magnetic charge $e$ are opposite to each other. The critical temperature and pressure  were  found highly sensitive to the variation parameters ( m and e).  


\begin{thebibliography}{99}
\bibitem{wh67}
 J. A. Wheeler, Our universe: the known and unknown," December, 1967. public lecture.
\bibitem{JD}
J.D. Bekenstein `` Black Hole and Entropy" Phy. Rev. D {\bf 7}, 2333 (1973).
\bibitem{JD1}
J.D. Bekenstein, `` Black Hole and the Second Law,"  Lett. al Nuovo Ciemnto 4 (1972) 15.
\bibitem{JD2} 
 J.D. Bekenstein `` Black Hole and Entropy" Phy. Rev. D {\bf 7}, 2333 (1973).
\bibitem{JD3}
 J.D. Bekenstein``Generialized Second Law of thermodynamics in  black hole Physics," Phy. Rev. D {\bf 9}, 3292 (1974).
\bibitem{JD4}
 J.D. Bekenstein``Statistical  black hole thermodynamics," Phy. Rev. D {\bf 12}, 3077 (1975). 
\bibitem{SWH} S.W. Hawking`` Black Hole and Thermodynamics" Phy. Rev. D {\bf 13}, 191 (1976).
\bibitem{jm}
J. M. Maldacena, Adv. Theor. Math. Phys. {\bf 2} 231 (1998). 



\bibitem{Regular:1968}
J. Bardeen~Regular,
in {\it Proceedings of GR5} (Tiflis, U.S.S.R., 1968).

\bibitem{AGB}   E.~Ayon-Beato and A.~Garcia,
Gen.\ Rel.\ Grav.\  {\bf 31}, 629 (1999).
\bibitem{AGB1}
 E. Ayon-Beato and A. Garcia, Gen. Rel. Grav.{\bf 37}, 635
(2005).
\bibitem{AGB2}E. Ayon-Beato, A. Garcia, Phys. Lett. B {\bf 493}, 149 (2000).
\bibitem{hc} H.~Culetu, arXiv:1408.3334v1 [gr-qc].
\bibitem{lbev}  L.~Balart and E.~C.~Vagenas,  Phys.\ Lett.\ B {\bf 730}, 14 (2014)
\bibitem{Balart:2014cga} L.~Balart and E.~C.~Vagenas, Phys.\ Rev.\ D {\bf 90}, no. 12, 124045 (2014).
\bibitem{Xiang} L.~Xiang, Y.~Ling and Y.~G.~Shen, Int.\ J.\ Mod.\ Phys.\ D {\bf 22}, 1342016 (2013).
	\bibitem{Bronnikov:2000vy}
	K.A.~Bronnikov,
	Phys.\ Rev.\ D {\bf 63}, 044005 (2001)
	
	\bibitem{Zaslavskii:2009kp}
	O.B.~Zaslavskii,
	Phys.\ Rev.\ D {\bf 80}, 064034 (2009)
	
	\bibitem{Lemos:2011dq}
	J.P.S.~Lemos and V.T.~Zanchin,
	Phys.\ Rev.\ D {\bf 83}, 124005 (2011)	.
\bibitem{dvs99}
D. V. Singh and S. Siwach, arXiv: 1909.11529 [hep-th].
\bibitem{Ansoldi:2008jw} 
S.~Ansoldi,
arXiv:0802.0330 [gr-qc].
\bibitem{Ghosh:2014pba} 
S.~G.~Ghosh,
Eur.\ Phys.\ J.\ C {\bf 75}, 532 (2015).
\bibitem{singh}
  D.~V.~Singh and N.~K.~Singh,
  Annals Phys.\  {\bf 383}, 600 (2017).

\bibitem{fr1}
S. Fernando, ” Int. Journal of Mod. Phys. D {\bf 26}, 1750071 (2017).

\bibitem{Bambi}  C.~Bambi and L.~Modesto,  Phys.\ Lett.\ B {\bf 721}, 329 (2013).

\bibitem{sharif}
M. Sharif, W. Javed, Can. J. Phys. 89, 1027 (2011).
\bibitem{sabir}
Md Sabir Ali, S.G. Ghosh, Phys. Rev. D 98, 084025 (2018)
\bibitem{dvs19}
D. V. Singh, S.G. Ghosh and S. D. Maharaj,  Annals Phys.\  {\bf 412}, 168025 (2019).
\bibitem{kumar19}
A. Kumar, D. V. Singh and S.G. Ghosh, Eur.\ Phys.\ J.\ C {\bf 79},  275 (2019).

\bibitem{Kumar:2020bqf}
A.~Kumar, D.~V.~Singh and S.~G.~Ghosh,
[arXiv:2003.14016 [gr-qc]].
\bibitem{Singh:2020xju}
D.~V.~Singh and S.~Siwach,
[arXiv:2003.11754 [gr-qc]].
\bibitem{Kumar:2020uyz}
A.~Kumar and R.~Kumar,
[arXiv:2003.13104 [gr-qc]].

\bibitem{Cai:2014znn}
  R.~G.~Cai, Y.~P.~Hu, Q.~Y.~Pan and Y.~L.~Zhang,
  Phys.\ Rev.\ D {\bf 91} (2015) no.2,  024032.
\bibitem{Babichev:2015xha}
E.~Babichev and R.~Brito,
Class. Quant. Grav. \textbf{32} (2015), 154001.
\bibitem{EslamPanah:2019fci}
B.~Eslam Panah and S.~Hendi,
EPL \textbf{125} (2019) no.6, 60006.
\bibitem{EslamPanah:2018rob}
B.~Eslam Panah, S.~Hendi and Y.~Ong,
Phys. Dark Univ. \textbf{27} (2020), 100452.


\bibitem{Tzikas:2018cvs}
A.~G.~Tzikas,
Phys. Lett. B \textbf{788} (2019), 219-224
\bibitem{Cvetic:2010jb}
  M.~Cvetic, G.~W.~Gibbons, D.~Kubiznak and C.~N.~Pope,
  Phys.\ Rev.\ D {\bf 84}  024037 (2011).
\bibitem{dolan13}
B. P. Dolan, D. Kastor, D. Kubiznak, R.  B. Mann, and J. Traschen, Phys.\ Rev.\ D {\bf 87}  104017 (2013).
\bibitem{hp}S. Hawking and D. Page, Commun. Math. Phys. {\bf 87}, 577 (1983).
\bibitem{witten}
E. Witten, Adv. Theor. Math. Phys. {\bf 2} 505 (1998).
\bibitem{Chamblin:1999hg}
  A.~Chamblin, R.~Emparan, C.~V.~Johnson and R.~C.~Myers,
  Phys.\ Rev.\ D {\bf 60} (1999) 104026.
\bibitem{Chamblin:1999tk}
  A.~Chamblin, R.~Emparan, C.~V.~Johnson and R.~C.~Myers,
  Phys.\ Rev.\ D {\bf 60} (1999) 06401.
\bibitem{1}
 D. Kubizˇnák, R.B. Mann, JHEP 1207, 033 (2012).
\bibitem{2}
 S.H. Hendi, M.H. Vahidinia, Phys. Rev. D 88, 084045 (2013).
\bibitem{3}
 R.-G. Cai, L.-M. Cao, L. Li, R.-Q. Yang, JHEP 1309, 005 (2013).
\bibitem{4}
 J.-X. Mo, W.-B. Liu, Phys. Lett. B 727, 336 (2013).
\bibitem{5}
 J.-X. Mo, W.-B. Liu, Eur. Phys. J. C 74, 2836 (2014).
\bibitem{6}
 J.-X. Mo, G.-Q. Li, W.-B. Liu, Phys. Lett. B 730, 111 (2014).
\bibitem{7}
 G.-Q. Li, Phys. Lett. B 735, 256 (2014).
\bibitem{8}
H.-H. Zhao, L.-C. Zhang, M.-S. Ma, R. Zhao, Phys. Rev. D 90,
064018 (2014).
\bibitem{9}
 M.H. Dehghani, S.Kamrani,A. Sheykhi, Phys. Rev.D90, 104020
(2014).
\bibitem{10}
 R.A. Hennigar,W.G. Brenna, R.B.Mann, JHEP 1507, 077 (2015).
\bibitem{11}
 J. Xu, L.M. Cao, Y.P. Hu, Phys. Rev. D 91, 124033 (2015).
\bibitem{12}
 S.H. Hendi, R.M. Tad, Z. Armanfard, M.S. Talezadeh, Eur. Phys.
J. C 76, 263 (2016).
\bibitem{13}
J. Sadeghi, Int. J. Theor. Phys. 55, 2455 (2016).
\bibitem{14}
J. Liang, C.-B. Sun, H.-T. Feng, Europhys.Lett. 113, 30008 (2016).
\bibitem{15}
 S. Fernando, Phys. Rev. D 94, 124049 (2016).
\bibitem{16}
 Z.-Y. Fan, Eur. Phys. J. C 77, 266 (2016).
\bibitem{17}
J. Sadeghi, B. Pourhassan, M. Rostami, Phys. Rev. D 94, 064006
(2016).
\bibitem{18}
 D. Hansen, D. Kubiznak, R.B. Mann, JHEP 1701, 047 (2017).
\bibitem{19}
 B.R. Majhi, S. Samanta, Phys. Lett. B 773, 203 (2017).
\bibitem{20}
 S.H. Hendi, B.E. Panah, S. Panahiyan,M.S. Talezadeh, Eur. Phys.
J. C 77, 133 (2017).
\bibitem{21}
 S. Upadhyay, B. Pourhassan, H. Farahani, Phys. Rev. D 95,
106014 (2017).
\bibitem{22}
 C.H. Nam, Eur. Phys. J. C 78, 581 (2018).
\bibitem{23}
 P. Pradhan, Mod. Phys. Lett. A 32, 1850030 (2018).
\bibitem{24}
D. Kastor and S Ray, Class. Quant. Grav. {\bf 26}, 195011 (2009). 

\bibitem{cao}
Cao H. Nam,  Eur.\ Phys.\ J.\ C {\bf 78}, 1016 (2018).
\bibitem{davis77}
P. Davis, Proc. R. Soc.A {\bf 353}, 499 (1977).


\end{thebibliography}
\end{document}